\documentclass[11pt,a4paper]{article}

\usepackage{amssymb}
\usepackage[dvips]{graphicx}
\usepackage{bm}

\usepackage{mathrsfs}
\usepackage{cite}

\begin{document}

\title{Finiteness of the two-loop matter contribution to the triple gauge-ghost vertices in ${\cal N}=1$ supersymmetric gauge theories regularized by higher derivatives}

\author{M.D.Kuzmichev, N.P.Meshcheriakov, S.V.Novgorodtsev, I.E.Shirokov, K.V.Stepanyantz
\medskip\\
{\small{\em Moscow State University, Faculty of Physics, Department of Theoretical Physics,}}\\
{\small{\em 119991, Moscow, Russia}}}

\maketitle

\begin{abstract}
For a general renormalizable ${\cal N}=1$ supersymmetric gauge theory with a simple gauge group we verify the ultraviolet (UV) finiteness of the two-loop matter contribution to the triple gauge-ghost vertices. These vertices have one leg of the quantum gauge superfield and two legs corresponding to the Faddeev--Popov ghost and antighost. By an explicit calculation made with the help of the higher covariant derivative regularization we demonstrate that the sum of the corresponding two-loop supergraphs containing a matter loop is not UV divergent in the case of using a general $\xi$-gauge. In the considered approximation this result confirms the recently proved theorem that the triple gauge-ghost vertices are UV finite in all orders, which is an important ingredient of the all-loop perturbative derivation of the Novikov-Shifman-Vainshtein-Zakharov relation.
\end{abstract}

\unitlength=1cm

\section{Introduction}
\hspace*{\parindent}

Possible ultraviolet divergences in supersymmetric theories are restricted by some non-renormalization theorems. For example, it is well known that the superpotential of ${\cal N}=1$ supersymmetric gauge theories cannot receive divergent quantum corrections \cite{Grisaru:1979wc}. Consequently, the renormalizations of masses and Yukawa couplings can be related to the renormalization of chiral matter superfields. However, there are also some other non-renormalization theorems even in theories with ${\cal N}=1$ supersymmetry. For example, it is reasonable to consider the exact Novikov-Shifman-Vainshtein-Zakharov (NSVZ) $\beta$-function \cite{Novikov:1983uc,Jones:1983ip,Novikov:1985rd,Shifman:1986zi} as a non-renormalization theorem, because it relates the renormalization of the gauge coupling constant to the renormalization of chiral matter superfields. Moreover, it produces the non-renormalization theorems for ${\cal N}=2$ \cite{Grisaru:1982zh,Howe:1983sr,Buchbinder:1997ib} and ${\cal N}=4$ \cite{Grisaru:1982zh,Howe:1983sr,Mandelstam:1982cb,Brink:1982pd} supersymmetric gauge theories \cite{Shifman:1999mv}. It is important that the non-renormalization theorems hold only for special renormalization prescriptions. Strictly speaking, even the non-renormalization of the superpotential requires either a manifestly supersymmetric superfield quantization or special limitations on a subtraction scheme. Therefore, it is highly desirable that the regularization and renormalization procedures be consistent with supersymmetry. Similarly, for deriving the finiteness of ${\cal N}=2$ supersymmetric gauge theories beyond the one-loop approximation from the NSVZ $\beta$-function one should use a manifestly ${\cal N}=2$ quantization procedure \cite{Buchbinder:2014wra}. Such a procedure can be constructed with the help of the harmonic superspace \cite{Galperin:1984av,Galperin:2001uw,Buchbinder:2001wy} and the corresponding invariant regularization \cite{Buchbinder:2015eva}. However, the NSVZ $\beta$-function is valid only for certain renormalization prescriptions, called ``the NSVZ schemes'', which constitute a continuous set \cite{Goriachuk:2018cac,Goriachuk:2020wyn}. It appeared that such popular renormalization schemes as $\overline{\mbox{DR}}$ and MOM do not enter this set, see Refs. \cite{Jack:1996vg,Jack:1996cn,Jack:1998uj,Harlander:2006xq,Mihaila:2013wma} and \cite{Kataev:2013csa,Kataev:2014gxa}, respectively. An all-loop prescription for constructing at least one of the NSVZ schemes was given in \cite{Stepanyantz:2020uke}.\footnote{In the Abelian case a similar prescription has been found earlier \cite{Kataev:2013eta} on the base of the results of \cite{Stepanyantz:2011jy,Stepanyantz:2014ima}.} The NSVZ scheme is obtained if a theory is regularized by the higher covariant derivative method \cite{Slavnov:1971aw,Slavnov:1972sq} (which includes introducing the Pauli--Villars determinants for removing one-loop divergences \cite{Slavnov:1977zf}) in the superfield version \cite{Krivoshchekov:1978xg,West:1985jx} and the renormalization is made by minimal subtractions of logarithms \cite{Kataev:2013eta}. This renormalization prescription is usually called HD+MSL \cite{Shakhmanov:2017wji,Stepanyantz:2017sqg}.\footnote{For ${\cal N}=1$ SQED the on-shell scheme appears to be another all-loop NSVZ renormalization prescription \cite{Kataev:2019olb}.} Note that actually the NSVZ $\beta$-function is valid in the HD+MSL scheme because it holds for RGFs defined in terms of the bare couplings for theories regularized by higher derivatives independently of a renormalization prescription. This statement has been verified by numerous multiloop calculations, see, e.g., \cite{Pimenov:2009hv,Stepanyantz:2012zz,Shakhmanov:2017soc,Kazantsev:2018nbl,Stepanyantz:2019lyo,Kuzmichev:2019ywn,Aleshin:2020gec}, and can be used for simple calculation of the $\beta$-function in higher orders \cite{Kazantsev:2020kfl}. The all-loop proof has been done in Refs. \cite{Stepanyantz:2020uke,Stepanyantz:2016gtk,Stepanyantz:2019ihw,Stepanyantz:2019lfm}. It turned out that for making this proof the NSVZ equation should be rewritten in an equivalent form \cite{Stepanyantz:2016gtk}, which does not contain the coupling constant dependent denominator similarly to the Abelian case \cite{Vainshtein:1986ja,Shifman:1985fi} and to the exact expression for the Adler $D$-function in ${\cal N}=1$ SQCD \cite{Shifman:2014cya,Shifman:2015doa}. The equivalence of both forms of the NSVZ relation can be established with the help of a non-renormalization theorem for the triple gauge-ghost vertices, which is an important ingredient needed for the perturbative proof of the exact NSVZ $\beta$-function. This theorem has been derived in \cite{Stepanyantz:2016gtk} for ${\cal N}=1$ supersymmetric gauge theories under the assumption of the superfield quantization in a general $\xi$-gauge. According to this theorem the triple gauge-ghost vertices in which one line corresponds to the {\it quantum} gauge superfield and two others correspond to the Faddeev--Popov ghost and antighost are UV finite in all orders. Earlier similar statements were known for theories formulated in terms of usual fields in the Landau gauge $\xi\to 0$ \cite{Dudal:2002pq,Capri:2014jqa}. They have been verified by three- and four-loop explicit calculations in Refs.  \cite{Capri:2014jqa} and \cite{Chetyrkin:2004mf}. In the general $\xi$-gauge the UV finiteness of the above mentioned vertices in the supersymmetric case was demonstrated by an explicit one-loop superfield calculation in Ref. \cite{Stepanyantz:2016gtk} made with the help of the higher covariant derivative regularization. In this paper we partially verify that this statement is also true in the two-loop approximation. Namely, we will prove that a part of the two-loop contribution to the triple gauge-ghost vertices coming from superdiagrams which contain a matter loop is UV finite for theories regularized by higher covariant derivatives. Note that we will use this regularization because it naturally produces the NSVZ scheme and reveals some interesting features of quantum corrections in supersymmetric theories, see \cite{Stepanyantz:2019lyo} and references therein. However, calculations of quantum corrections with this regularization are rather complicated and to a certain degree are similar to the ones for higher derivative theories (see, e.g., \cite{BezerradeMello:2016bjn,Gama:2017ets,Gama:2020pte}).

The paper is organized as follows. In Sect. \ref{Section_SUSY_Theories} we recall the superfield formulation of ${\cal N}=1$ supersymmetric gauge theories together with some aspects of their regularization by higher derivatives and superfield quantization. The structure of the triple gauge-ghost vertices is discussed in Sect. \ref{Section_Vertices}. The calculation of the two-loop superdiagrams containing a matter loop is described in Sect. \ref{Section_Matter_Contribution}, where we prove that their overall contribution is not UV divergent.

\section{${\cal N}=1$ supersymmetric gauge theories and the regularization by higher covariant derivatives}
\hspace{\parindent}\label{Section_SUSY_Theories}

We will consider a general renormalizable ${\cal N}=1$ supersymmetric gauge theory with a single gauge coupling constant. In the superfield formulation its classical action is written in the form

\begin{eqnarray}\label{Action_Classical_Superfield}
&& S = \frac{1}{2 e_0^2}\,\mbox{Re}\,\mbox{tr}\int d^4x\,
d^2\theta\,W^a W_a + \frac{1}{4} \int d^4x\, d^4\theta\,\phi^{*i}
(e^{2V})_i{}^j \phi_j\nonumber\\
&&\qquad\qquad\qquad\qquad\qquad\qquad  +
\bigg\{\int d^4x\,d^2\theta\,\Big(\frac{1}{4} m_0^{ij} \phi_i \phi_j + \frac{1}{6}\lambda_0^{ijk} \phi_i
\phi_j \phi_k\Big) + \mbox{c.c.}\bigg\},\qquad
\end{eqnarray}

\noindent
where $e_0$ and $\lambda_0^{ijk}$ are the bare gauge and Yukawa couplings, respectively, and $m_0^{ij}$ is the bare mass matrix. The Hermitian gauge superfield and its superfield strength are denoted by $V$ and $W_a$, respectively. The chiral matter superfields $\phi_i$ lie in a certain representation $R$ of the gauge group $G$. In our notations the generators of the fundamental representation denoted by $t^A$ are normalized by the condition $\mbox{tr}(t^A t^B) = \delta^{AB}/2$, while the generators of the representation $R$ are denoted by $T^A$ and satisfy the equations

\begin{equation}
\mbox{tr}(T^A T^B) = T(R) \delta^{AB};\qquad (T^A T^A)_i{}^j = C(R)_i{}^j.
\end{equation}

\noindent
The theory is gauge invariant if the bare masses and Yukawa couplings are chosen in such a way that

\begin{eqnarray}\label{Mass_Invariance}
&& m_0^{ik} (T^A)_k{}^{j} + m_0^{kj} (T^A)_k{}^{i} = 0;\vphantom{\Big(}\\
\label{Yukawa_Invariance}
&& \lambda_0^{ijm} (T^A)_m{}^{k} + \lambda_0^{imk} (T^A)_m{}^{j} +
\lambda_0^{mjk} (T^A)_m{}^{i} = 0.\vphantom{\Big(}
\end{eqnarray}

\noindent
Below we will always assume that these equations are satisfied. Also we will always assume that

\begin{equation}\label{Mass_Squared}
m_0^{ik} m^*_{0kj} = m_0^2 \delta_j^i.
\end{equation}

Note that these conditions can be satisfied only for anomaly free theories. Really, using Eqs. (\ref{Mass_Invariance}) and (\ref{Mass_Squared}) after some transformations we obtain

\begin{eqnarray}
&& m_0^2\, \mbox{tr} (T^A T^B T^C) = m^*_{0ij} m_0^{jk} (T^A)_k{}^l (T^B)_l{}^m (T^C)_m{}^i \vphantom{\Big(}\nonumber\\
&&\qquad\qquad = - m^*_{0ij} m_0^{mi} (T^A)_k{}^j (T^B)_l{}^k (T^C)_m{}^l = - m_0^2\, \mbox{tr}(T^A T^C T^B). \qquad\vphantom{\Big(}
\end{eqnarray}

\noindent
This implies that the generators $T^A$ should satisfy the anomaly cancellation condition \cite{Bertlmann:1996xk}

\begin{equation}\label{Anomalies}
\mbox{tr}(T^A \{T^B, T^C\}) = 0.
\end{equation}

\noindent
Certainly, the absence of the gauge anomalies is also needed for the renormalizability, which will essentially be used in what follows.

For quantizing the theory (\ref{Action_Classical_Superfield}) it is convenient to use the background field method. Moreover, one should take into account that the quantum gauge superfield is renormalized in a nonlinear way \cite{Piguet:1981fb,Piguet:1981hh,Tyutin:1983rg}. This has been confirmed by explicit calculations in the lowest orders of the perturbation theory \cite{Juer:1982fb,Juer:1982mp}. Also explicit calculations demonstrate that without the nonlinear renormalization the renormalization group equations are not satisfied \cite{Kazantsev:2018kjx}. To take into account the nonlinear renormalization and to introduce the quantum-background splitting, we make the substitution

\begin{equation}\label{Splitting}
e^{2V} \to e^{2{\cal F}(V)} e^{2\bm{V}}.
\end{equation}

\noindent
Here $V$ and $\bm{V}$ denote the quantum and background gauge superfields, respectively. Note that in this notation the quantum gauge superfield satisfies the equation $V^+ = e^{-2\bm{V}} V e^{2\bm{V}}$. After the replacement (\ref{Splitting}) the gauge superfield strength takes the form

\begin{equation}
W_a = \frac{1}{8} \bar D^2 \left(e^{-2\bm{V}} e^{-2{\cal F}(V)} D_a \left(e^{2{\cal F}(V)} e^{2\bm{V}}\right)\right).
\end{equation}

In this paper we will consider only superdiagrams which do not contain external lines of the background superfield. However, for other purposes the background (super)field method is very useful, so that constructing the generating functional we will keep the dependence on the background gauge superfield $\bm{V}$.

Following Refs. \cite{Aleshin:2016yvj,Kazantsev:2017fdc}, we introduce the regularization by adding some terms containing higher derivatives to the action. After this the regularized action can be written in the form

\begin{eqnarray}\label{Action_Regularized}
&& S_{\mbox{\scriptsize reg}} = \frac{1}{2 e_0^2}\,\mbox{Re}\, \mbox{tr} \int d^4x\, d^2\theta\, W^a \Big[e^{-2\bm{V}} e^{-2{\cal F}(V)}\,  R\Big(-\frac{\bar\nabla^2 \nabla^2}{16\Lambda^2}\Big)\, e^{2{\cal F}(V)}e^{2\bm{V}}\Big]_{Adj} W_a \qquad\nonumber\\
&& + \frac{1}{4} \int d^4x\,d^4\theta\, \phi^{*i} \Big[\, F\Big(-\frac{\bar\nabla^2 \nabla^2}{16\Lambda^2}\Big) e^{2{\cal F}(V)}e^{2\bm{V}}\Big]_i{}^j \phi_j
+ \bigg\{ \int d^4x\,d^2\theta\, \Big(\frac{1}{4} m_0^{ij} \phi_i \phi_j \qquad\nonumber\\
&& + \frac{1}{6} \lambda_0^{ijk} \phi_i \phi_j \phi_k\Big) + \mbox{c.c.} \bigg\},\qquad
\end{eqnarray}

\noindent
where $\Lambda$ is the dimensionful cut-off parameter of the regularized theory, and the covariant derivatives are defined as

\begin{equation}\label{Covariant_Derivative_Definition}
\nabla_a = D_a;\qquad \bar\nabla_{\dot a} = e^{2{\cal F}(V)} e^{2\bm{V}} \bar D_{\dot a} e^{-2\bm{V}} e^{-2{\cal F}(V)}.
\end{equation}

\noindent
(The higher derivatives are present inside two regulator functions $R(x)$ and $F(x)$, which rapidly grow at infinity and are equal to 1 at $x=0$.) In our notations, if $f(x) = f_0 + f_1 x + f_2 x^2 + \ldots$,
then the subscript $Adj$ means that

\begin{equation}
f(X)_{Adj} Y \equiv f_0 Y + f_1 [X,Y] + f_2 [X,[X,Y]] +\ldots
\end{equation}

The gauge fixing term analogous to the background $\xi$-gauge in the usual Yang--Mills theory is given by the expression

\begin{equation}\label{Term_For_Fixing_Gauge}
S_{\mbox{\scriptsize gf}} = -\frac{1}{16\xi_0 e_0^2}\, \mbox{tr} \int d^4x\, d^4\theta\,  \bm{\nabla}^2 V K\Big(-\frac{\bm{\bar\nabla}^2 \bm{\nabla}^2}{16\Lambda^2}\Big)_{Adj} \bm{\bar\nabla}^2 V,
\end{equation}

\noindent
which includes the background covariant derivatives $\bm{\nabla}_a \equiv D_a$ and $\bm{\bar\nabla}_{\dot a} \equiv e^{2\bm{V}} \bar D_{\dot a} e^{-2\bm{V}}$. Also the gauge fixing term contains one more higher derivative regulator function $K(x)$, which has the same properties as the functions $R(x)$ and $F(x)$. Then the actions for the chiral Faddeev--Popov ghosts $c$ and $\bar c$ and the chiral Nielsen--Kallosh ghosts $b$ read

\begin{eqnarray}
\label{Ghosts_Faddeev-Popov_Action}
&& S_{\mbox{\scriptsize FP}} = \frac{1}{2} \int
d^4x\,d^4\theta\, \frac{\partial {\cal F}^{-1}(\widetilde V)^A}{\partial {\widetilde V}^B}\left.\vphantom{\frac{1}{2}}\right|_{\widetilde V = {\cal F}(V)} \left(e^{2\bm{V}}\bar c e^{-2\bm{V}} +
\bar c^+ \right)^A\nonumber\\
&&\qquad\qquad\qquad\quad \times \left\{\vphantom{\frac{1}{2}} \smash{\Big(\frac{{\cal F}(V)}{1-e^{2{\cal F}(V)}}\Big)_{Adj} c^+
+ \Big(\frac{{\cal F}(V)}{1-e^{-2{\cal F}(V)}}\Big)_{Adj}
\Big(e^{2\bm{V}} c e^{-2\bm{V}}\Big)}\right\}^B;\qquad\\
&& \vphantom{1}\nonumber\\
\label{Ghosts_Nielsen-Kallosh_Action}
&&  S_{\mbox{\scriptsize NK}} = \frac{1}{2e_0^2}\,\mbox{tr} \int d^4x\,d^4\theta\, b^+ \Big(K\Big(-\frac{\bm{\bar\nabla}^2 \bm{\nabla}^2}{16\Lambda^2}\Big) e^{2\bm{V}}\Big)_{Adj} b.
\end{eqnarray}

To regularize one-loop divergences that survive after introducing the higher derivatives, we should insert the Pauli--Villars determinants into the generating functional \cite{Slavnov:1977zf}. According to \cite{Aleshin:2016yvj,Kazantsev:2017fdc}, in the supersymmetric case one needs two such determinants. The first one can be presented as a functional integral over three commuting chiral superfields $\varphi_1$, $\varphi_2$, and $\varphi_3$ in the adjoint representation of the gauge group,

\begin{equation}
\mbox{Det}(PV,M_\varphi)^{-1} = \int D\varphi_1 D\varphi_2 D\varphi_3 \exp(i S_\varphi),
\end{equation}

\noindent
where

\begin{eqnarray}
&& S_\varphi = \frac{1}{2e_0^2} \mbox{tr}\int d^4x\, d^4\theta\, \Big(\varphi_1 ^+ \Big[ R\Big(-\frac{\bar\nabla^2 \nabla^2}{16\Lambda^2}\Big)e^{2{\cal F}(V)}  e^{2\bm{V}}\Big]_{Adj}\varphi_1 + \varphi_2^+ \Big[ e^{2{\cal F}(V)} e^{2\bm{V}}\Big]_{Adj}\varphi_2\qquad\nonumber\\
&& + \varphi_3^+ \Big[ e^{2{\cal F}(V)} e^{2\bm{V}}\Big]_{Adj}\varphi_3\Big) + \frac{1}{2e_0^2}\Big(\mbox{tr}\int d^4x\, d^2\theta\, M_\varphi (\varphi_1^2 + \varphi_2^2 + \varphi_3^2) +\mbox{c.c.}\Big).\qquad
\end{eqnarray}

\noindent
This determinant cancels one-loop divergences generated by the gauge and ghost superfields. The second Pauli--Villars determinant removes one-loop divergences produced by a matter loop. It is given by the functional integral over the (commuting) chiral superfields $\Phi_i$ in a representation $R_{\mbox{\scriptsize PV}}$ which admits a gauge invariant mass term such that $M^{ik} M^*_{kj} = M^2 \delta_j^i$,

\begin{equation}
\mbox{Det}(PV,M)^{-1} = \int D\Phi \exp(iS_\Phi),
\end{equation}

\noindent
where

\begin{equation}
S_\Phi = \frac{1}{4} \int d^4x\, d^4\theta\, \Phi^+ F\Big(-\frac{\bar\nabla^2 \nabla^2}{16\Lambda^2}\Big)e^{2{\cal F}(V)} e^{2\bm{V}} \Phi
+ \Big(\frac{1}{4}\int d^4x\, d^2\theta\, M^{ij} \Phi_i \Phi_j +\mbox{c.c.}\Big).
\end{equation}

\noindent
Then the generating functional of the regularized theory takes the form

\begin{eqnarray}\label{Generating_Functional_Z}
&& Z[\bm{V},\mbox{Sources}] = \int D\mu\,
\Big(\mbox{Det}(PV,M)\Big)^c \mbox{Det}(PV,M_{\varphi})^{-1}
\qquad\nonumber\\
&&\qquad\qquad\qquad\qquad\qquad \times \exp\Big(iS_{\mbox{\scriptsize reg}} +
iS_{\mbox{\scriptsize gf}} + i S_{\mbox{\scriptsize FP}} + i
S_{\mbox{\scriptsize NK}} + i S_{\mbox{\scriptsize
sources}}\Big),\qquad
\end{eqnarray}

\noindent
where $D\mu$ denotes the integration measure, $c = T(R)/T(R_{\mbox{\scriptsize PV}})$ and $S_{\mbox{\scriptsize sources}}$ includes all relevant sources. Moreover, to obtain a theory with a single dimensionful regularization parameter, we require that the ratios $a_\varphi\equiv M_\varphi/\Lambda$ and $a\equiv M/\Lambda$ are constants which do not depend on couplings. The effective action $\Gamma$ is constructed according to the standard procedure, as a Legendre transform of the generating functional for the connected Green functions $W\equiv -i\ln Z$.

It is important that  both the regularized action and the above described Pauli--Villars determinants are gauge invariant. Due to the use of the background field method the original gauge invariance produces two different types of transformations. The background gauge symmetry

\begin{eqnarray}
&&\hspace*{-5mm} \phi_i \to (e^A)_i{}^j \phi_j;\qquad V \to e^{-A^+} V e^{A^+};\qquad e^{2\bm{V}} \to e^{-A^+} e^{2\bm{V}} e^{-A}; \qquad \varphi_{1,2,3} \to e^A \varphi_{1,2,3} e^{-A}; \vphantom{\Big(} \nonumber\\
&&\hspace*{-5mm}\qquad\qquad \Phi_i \to (e^A)_i{}^j \Phi_j;\qquad c \to e^A c e^{-A};\qquad \bar c\to e^A \bar c e^{-A};\qquad b \to e^A b e^{-A} \vphantom{\Big(}
\end{eqnarray}

\noindent
parameterized by a chiral Lie-algebra valued superfield $A$ remains unbroken and is a manifest symmetry of the effective action. In contrast, the quantum gauge invariance is broken by the gauge fixing procedure. However, the remaining BRST symmetry produces the Slavnov--Taylor identities, which can be derived with the help of the standard procedure \cite{Taylor:1971ff,Slavnov:1972fg}. The regularized theory described by the generating functional (\ref{Generating_Functional_Z}) is finite (for finite value of $\Lambda$) and gauge invariant. Therefore, no anomalies can appear {\it in the gauge Slavnov--Taylor identities} due to the absence of ambiguous linearly divergent integrals.\footnote{Even in usual QED the Pauli--Villars regularization allows calculating the anomaly of the axial current in such a way that the gauge symmetry is automatically unbroken, see, e.g., \cite{Bertlmann:1996xk}.} However, it is well known that in general the anomalies can appear in the gauge Slavnov--Taylor identities. The contradiction is solved if we take into account that one can introduce the considered regularization (which ensures the absence of gauge anomalies) only if Eq. (\ref{Anomalies}) and the similar condition for the generators of the representation $R_{\mbox{\scriptsize PV}}$

\begin{equation}\label{Anomalies_PV}
\mbox{tr}\Big(T^A_{\mbox{\scriptsize PV}}\{T^B_{\mbox{\scriptsize PV}},T^C_{\mbox{\scriptsize PV}}\}\Big) = 0
\end{equation}

\noindent
are satisfied.\footnote{Evidently, the analogous equation is always valid for the adjoint representation.} Otherwise, the gauge symmetry is broken by the mass terms. However, the gauge anomalies are proportional to the structure (\ref{Anomalies}) even in the case of using ${\cal N}=1$ superfield quantization \cite{Ohshima:1999jg}. This implies that the considered version of the regularization can be constructed only if the gauge anomalies are absent, so that no contradiction appears.

\section{Structure of the three-point gauge-ghost vertices and their finiteness}
\hspace*{\parindent}\label{Section_Vertices}

We are interested in the 3-point vertices with two external ghost legs and one external leg of the quantum gauge superfield. (Note that similar vertices with a leg of the background gauge superfield are in general UV divergent.) There are four different vertices of the considered structure, namely, $\bar c^+ V c$, $\bar c V c$, $\bar c^+ V c^+$, and $\bar c V c^+$ depending on the (anti)ghost superfields on the external lines. According to \cite{Stepanyantz:2016gtk} all these vertices have the same renormalization constant $Z_\alpha^{-1/2} Z_c Z_V$, where $Z_\alpha$, $Z_c$, and $Z_V$ are the renormalization constants for the gauge coupling constant $\alpha = e^2/4\pi$, the Faddeev--Popov ghosts, and the quantum gauge superfield, respectively,

\begin{equation}\label{Renormalization_Constants}
\frac{1}{\alpha_0} = \frac{Z_\alpha}{\alpha};\qquad \bar c^A c^B = Z_c\, \bar c_R^A c_R^B; \qquad V^A = Z_V V_R^A,
\end{equation}

\noindent
where the subscript $R$ denotes renormalized superfields. Certainly, it should be noted that the quantum gauge superfield is renormalized in a nonlinear way. To take this nonlinear renormalization into account, we include an infinite set of parameters into the function ${\cal F}(V)^A$. Say, in the lowest nontrivial order it is given by the expression \cite{Juer:1982fb,Juer:1982mp}

\begin{equation}\label{Function_F}
{\cal F}(V)^A = V^A + e_0^2\, y_0\, G^{ABCD} V^B V^C V^D + \ldots,
\end{equation}

\noindent
where $G^{ABCD} \equiv \big(f^{AKL} f^{BLM} f^{CMN} f^{DNK} + \mbox{permutations of $B$, $C$, and $D$}\big)/6$, and contains a parameter $y_0$, which should also be renormalized. Then the nonlinear renormalization is reduced to linear renormalizations of $V^A$ and of the parameters $y_0, \ldots$ The equation describing the renormalization of the parameter $y_0$ in the lowest nontrivial approximation can be found, e.g., in \cite{Kazantsev:2018kjx}.

The structure of the triple gauge-ghost vertices can be analysed with the help of dimensional and chirality considerations. Using notations similar to the ones in Ref. \cite{Stepanyantz:2016gtk} we write the corresponding parts of the effective action as

\begin{eqnarray}\label{Three_Point_Contribution1}
&&\hspace*{-5mm} \Delta\Gamma_{\bar c^+ V c} = \frac{i e_0}{4} f^{ABC} \int d^4\theta\, \frac{d^4p}{(2\pi)^4} \frac{d^4q}{(2\pi)^4} \bar c^{+A}(p+q,\theta)\Big(s(p,q) \partial^2\Pi_{1/2}V^B(-p,\theta)\nonumber\\
&&\hspace*{-5mm} \qquad\qquad\qquad\quad\ \ \, + {\cal S}_\mu(p,q) (\gamma^\mu)_{\dot a}{}^{b} D_b \bar D^{\dot a} V^B(-p,\theta) + {\cal S}(p,q) V^B(-p,\theta)\Big) c^C(-q,\theta);\qquad\\
\label{Three_Point_Contribution2}
&&\hspace*{-5mm} \Delta\Gamma_{\bar c^+ V c^+} = \frac{i e_0}{4} f^{ABC} \int d^4\theta\, \frac{d^4p}{(2\pi)^4} \frac{d^4q}{(2\pi)^4} \bar c^{+A}(p+q,\theta) \widetilde {\cal S}(p,q) V^B(-p,\theta) c^{+C}(-q,\theta),\qquad
\end{eqnarray}

\noindent
where $\partial^2\Pi_{1/2} \equiv - D^a \bar D^2 D_a/8$ is a supersymmetric analog of the transversal projection operator. Differentiating these expressions with respect to superfields we present the considered Green functions in the form

\begin{eqnarray}\label{Gauge_Ghost1}
&&\hspace*{-5mm} \frac{\delta^3\Gamma}{\delta \bar c_x^{+A} \delta V_y^B \delta c_z^C} = -\frac{i e_0}{16} f^{ABC} \int \frac{d^4p}{(2\pi)^4} \frac{d^4q}{(2\pi)^4} \Big(s(p,q) \partial^2\Pi_{1/2} \nonumber\\
&&\hspace*{-5mm}\qquad\qquad\quad\ \ \, - {\cal S}_\mu(p,q) (\gamma^\mu)_{\dot a}{}^{b} \bar D^{\dot a} D_b + {\cal S}(p,q) \Big)_{y} \Big(D_{x}^2\delta^8_{xy}(q+p)\, \bar D_{z}^2 \delta^8_{yz}(q)\Big);\qquad\\
&&\vphantom{1} \nonumber\\
\label{Gauge_Ghost2}
&&\hspace*{-5mm}  \frac{\delta^3\Gamma}{\delta \bar c_x^{+A} \delta V_y^B \delta c_z^{+C}} = -\frac{i e_0}{16} f^{ABC} \int \frac{d^4p}{(2\pi)^4} \frac{d^4q}{(2\pi)^4} \widetilde {\cal S}(p,q) D_x^2\delta^8_{xy}(q+p) D_z^2 \delta^8_{yz}(q),
\end{eqnarray}

\noindent
where $\delta^8_{xy}(q) \equiv e^{iq_\mu (x^\mu-y^\mu)} \delta^4(\theta_x-\theta_y)$.

The one-loop expressions for the functions $s$, ${\cal S}_\mu$, ${\cal S}$, and $\widetilde {\cal S}$ can be found in Ref. \cite{Stepanyantz:2016gtk} (in which they are denoted by $f$, $F_\mu$, $F$, and $\widetilde F$, respectively). For example, the sum of the tree and one-loop contributions to ${\cal S}$ is given by the following function of the Euclidean momenta $P$ and $Q$\footnote{In our conventions Euclidean momenta are always denoted by capital letters.}

\begin{eqnarray}
&&\hspace*{-7mm} {\cal S}(P,Q) = 1 + \frac{e_0^2 C_2}{4} \int \frac{d^4K}{(2\pi)^4} \Bigg\{-\frac{(Q+P)^2}{R_K K^2 (K+P)^2 (K-Q)^2} - \frac{\xi_0\, P^2}{K_K K^2 (K+Q)^2 (K+Q+P)^2}\nonumber\\
&&\hspace*{-7mm} + \frac{\xi_0\, Q^2}{K_K K^2 (K+P)^2 (K+Q+P)^2} + \left(\frac{\xi_0}{K_K} - \frac{1}{R_K}\right)\left(- \frac{2(Q+P)^2}{K^4 (K+Q+P)^2}  + \frac{2}{K^2 (K+Q+P)^2} \right.\nonumber\\
&&\hspace*{-7mm} \left. -\frac{1}{K^2 (K+Q)^2} - \frac{1}{K^2 (K+P)^2} \right)\Bigg\}+ O(\alpha_0^2,\alpha_0\lambda_0^2),
\end{eqnarray}

\noindent
where $R_K \equiv R(K^2/\Lambda^2)$ etc. We see that this expression is finite in the UV region independently of the value of the gauge parameter $\xi_0$, although some terms inside it are logarithmically divergent. The function $\widetilde {\cal S}$ is given by a similar UV finite expression. In the one-loop approximation the UV finiteness of the functions ${\cal S}_\mu$ and $s$ immediately follows from the fact that they have the dimensions $m^{-1}$ and $m^{-2}$, respectively.

In this paper we would like to verify that a part of the two-loop contribution to the Green functions (\ref{Gauge_Ghost1}) and (\ref{Gauge_Ghost2}) coming from supergraphs containing a matter loop is UV finite. The straightforward calculation is rather complicated, especially due to the use of the regularization by higher covariant derivatives. However, it is possible to make some simplifications. First, we know that all 4 three-point gauge-ghost vertices have the same renormalization constants. Therefore, it is sufficient to calculate only one of them. In this paper we will consider the function (\ref{Gauge_Ghost1}). Moreover, the integrals giving the functions ${\cal S}_\mu$ and $s$ have the superficial degree of divergence $-1$ and $-2$, respectively. This implies that the corresponding divergences can come only from divergent subdiagrams. For the considered renormalizable theory these subdivergences are evidently removed by the renormalization in the previous orders. Therefore, to find the two-loop contribution to the renormalization constant $Z_\alpha^{-1/2} Z_c Z_V$, we need to calculate only the function ${\cal S}$. This function can be extracted from the corresponding part of the effective action (given by Eq. (\ref{Three_Point_Contribution1})) by a formal substitution\footnote{This substitution is needed only for extracting a certain part of the Green function, so that it is not essential that the expression $\bar D^2 H$ is not Hermitian.}

\begin{equation}\label{Formal_Substitution}
V\to \bar D^2 H,
\end{equation}

\noindent
where $H$ is a Hermitian superfield, because after this substitution the expression (\ref{Three_Point_Contribution1}) takes the form

\begin{equation}
\Delta\Gamma_{\bar c^+ V c} = \frac{i e_0}{4} f^{ABC} \int d^4\theta\, \frac{d^4p}{(2\pi)^4} \frac{d^4q}{(2\pi)^4} \bar c^{+A}(p+q,\theta) {\cal S}(p,q) \bar D^2 H^B(-p,\theta) c^C(-q,\theta).
\end{equation}

\noindent
Moreover, the calculation of the function ${\cal S}$ can be done in the limit of the vanishing external momenta. Really, terms proportional to external momenta are given by integrals with a negative superficial degree of divergence, so that after removing subdivergences by the renormalization in the previous orders we will obtain UV finite contributions.

Thus, we will extract the function ${\cal S}$ with the help of the formal substitution (\ref{Formal_Substitution}) and calculate it in the limit of the vanishing external momenta, $P,Q\to 0$. Details of this calculation we describe in the next section.

\section{The matter contribution to the triple gauge-ghost vertices}
\hspace*{\parindent}\label{Section_Matter_Contribution}

\begin{figure}[h]
\begin{picture}(0,11.5)
\put(1,9){\includegraphics[scale=0.2,clip]{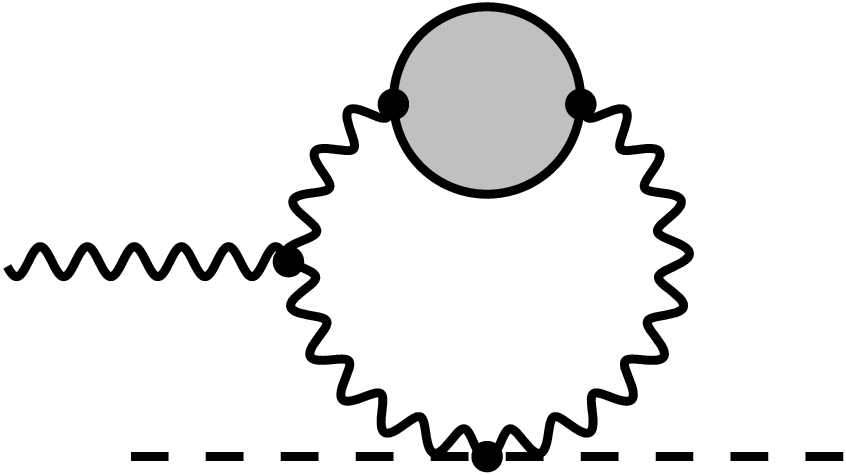}}
\put(1,11){(1)} \put(1.1,8.7){$\bar c^+$} \put(3.8,8.7){$c$}
\put(5,9){\includegraphics[scale=0.2,clip]{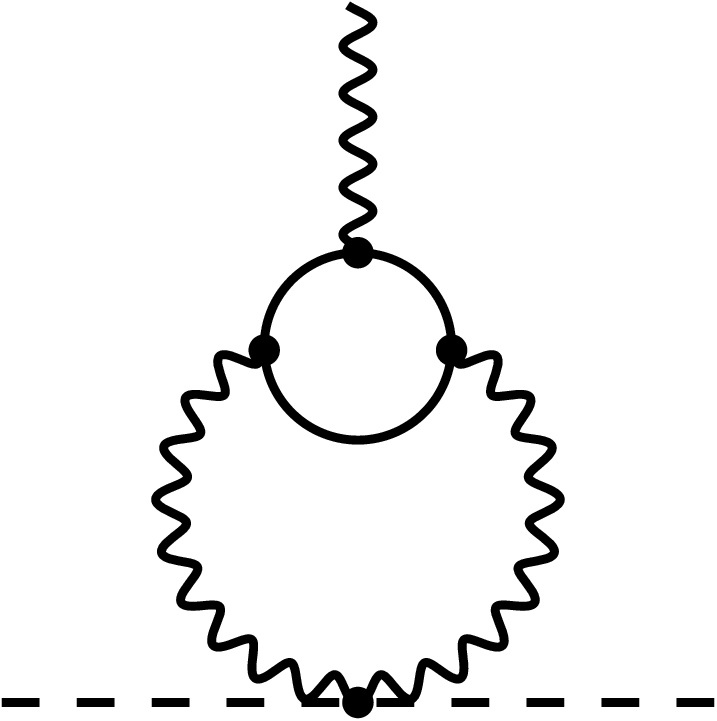}}
\put(5,11){(2)}
\put(9,9){\includegraphics[scale=0.2,clip]{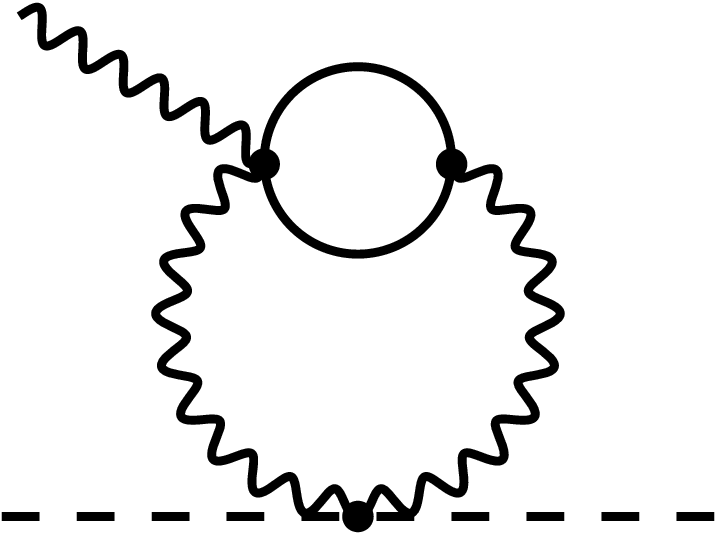}}
\put(9,11){(3)}
\put(13,9){\includegraphics[scale=0.2,clip]{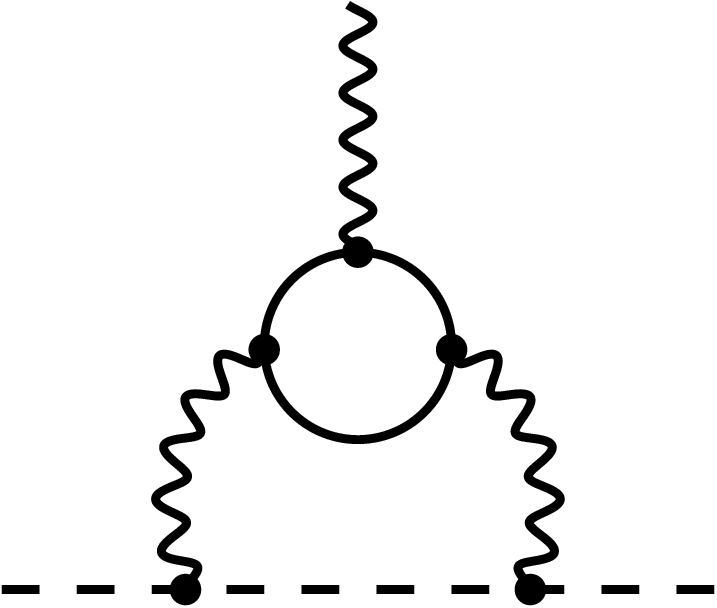}}
\put(13,11){(4)}

\put(1,6.2){\includegraphics[scale=0.2,clip]{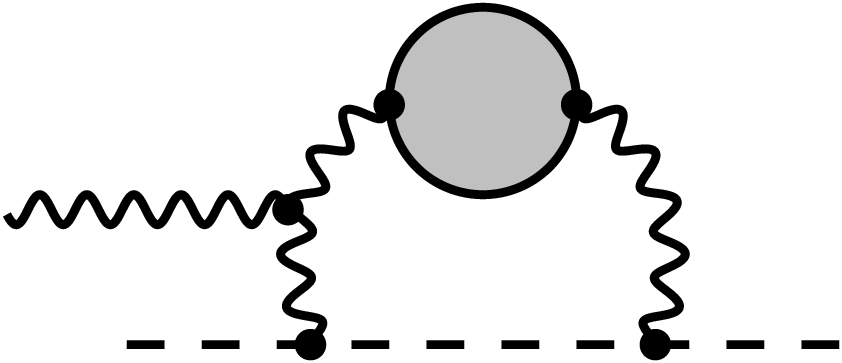}}
\put(1,7.9){(5)}
\put(5,6.2){\includegraphics[scale=0.2,clip]{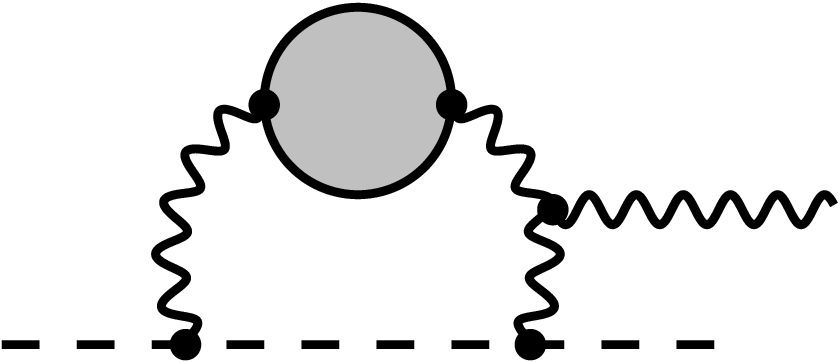}}
\put(5,7.9){(6)}
\put(9,6.2){\includegraphics[scale=0.2,clip]{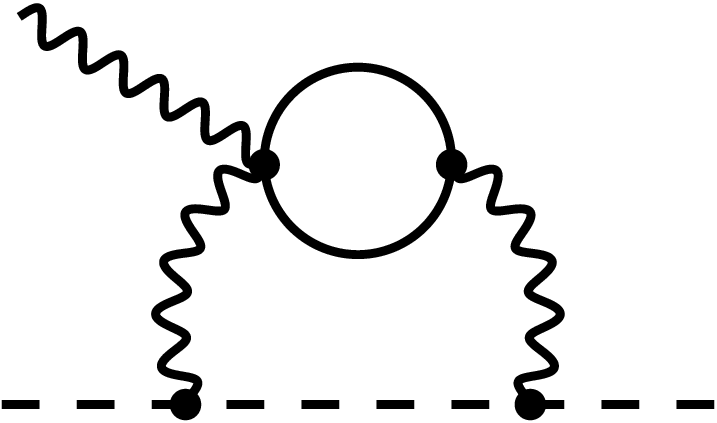}}
\put(9,7.9){(7)}
\put(13,6.2){\includegraphics[scale=0.2,clip]{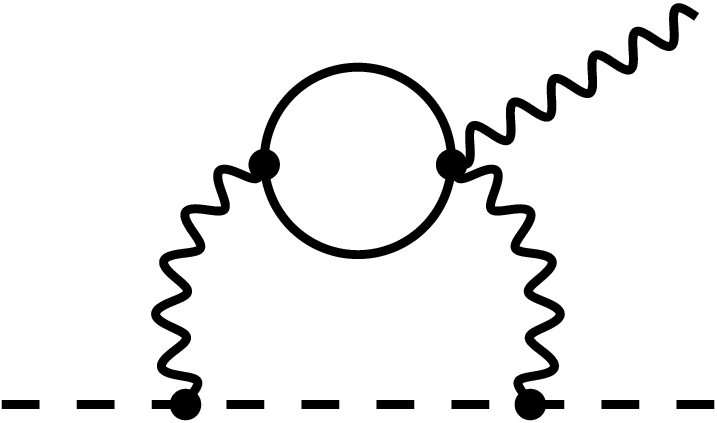}}
\put(13,7.9){(8)}

\put(1,3.1){\includegraphics[scale=0.2,clip]{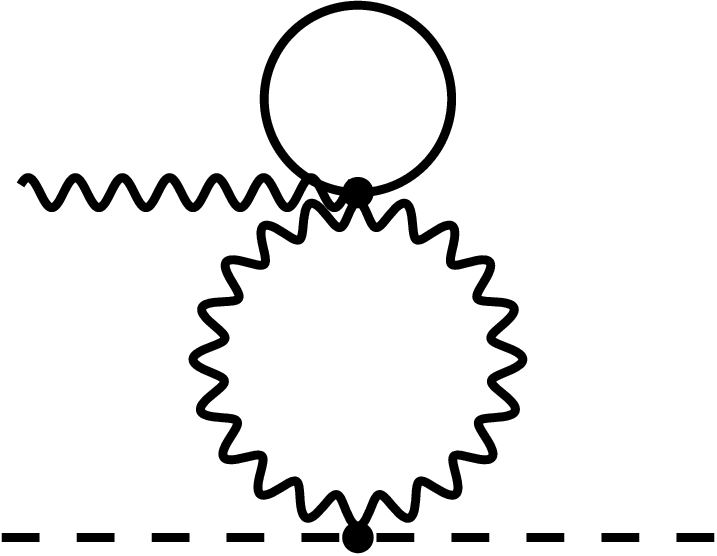}}
\put(1,5.1){(9)}
\put(5,3.1){\includegraphics[scale=0.2,clip]{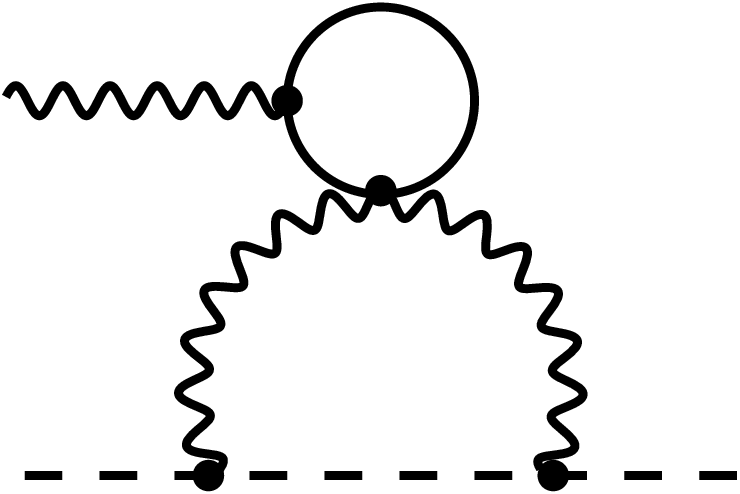}}
\put(5,5.1){(10)}
\put(9,3.1){\includegraphics[scale=0.2,clip]{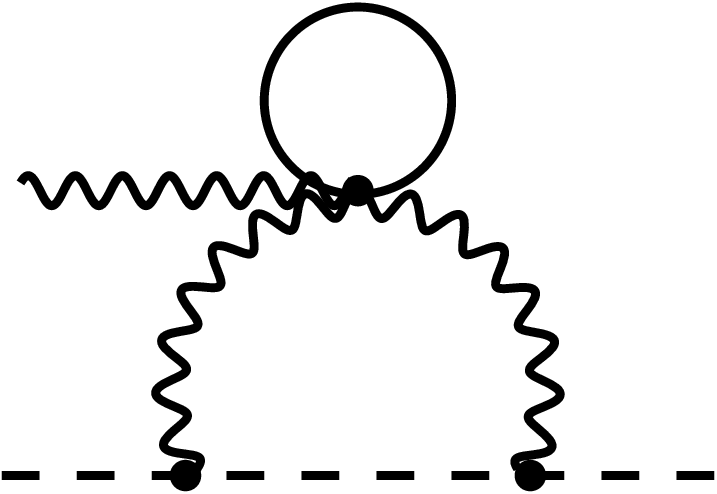}}
\put(9,5.1){(11)}
\put(13,3.1){\includegraphics[scale=0.2,clip]{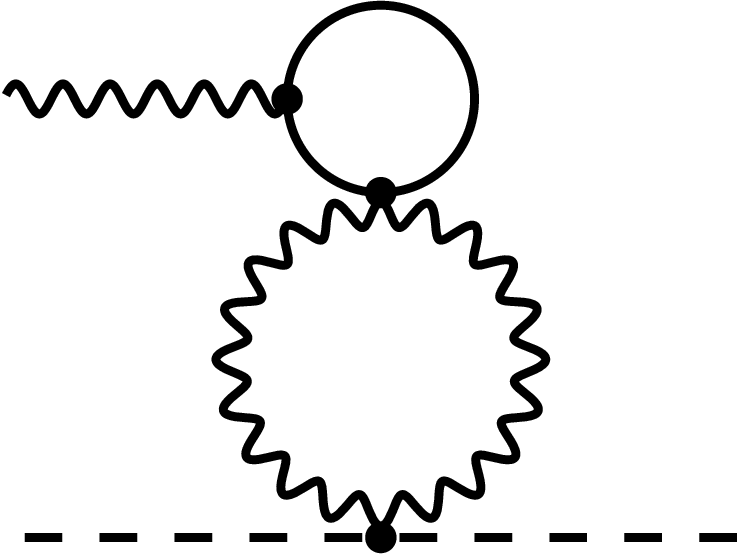}}
\put(13,5.1){(12)}

\put(3.5,-0.1){\includegraphics[scale=0.2,clip]{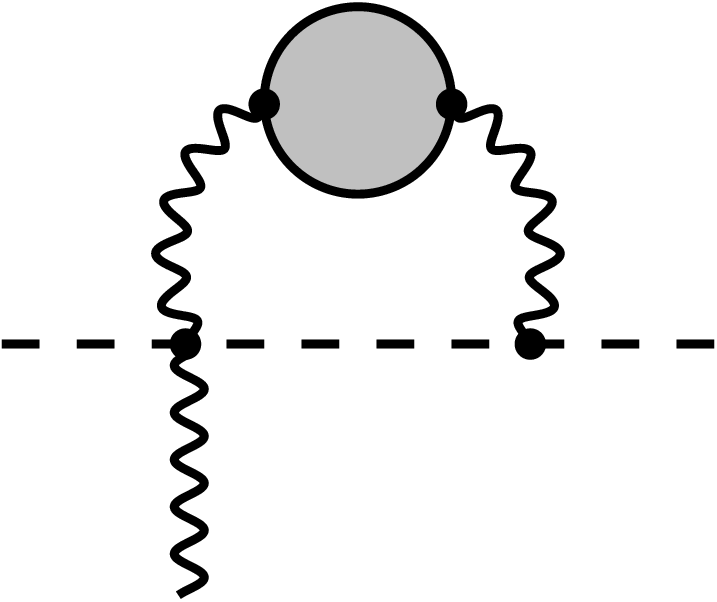}}
\put(3.5,1.9){(13)}
\put(7.5,-0.1){\includegraphics[scale=0.2,clip]{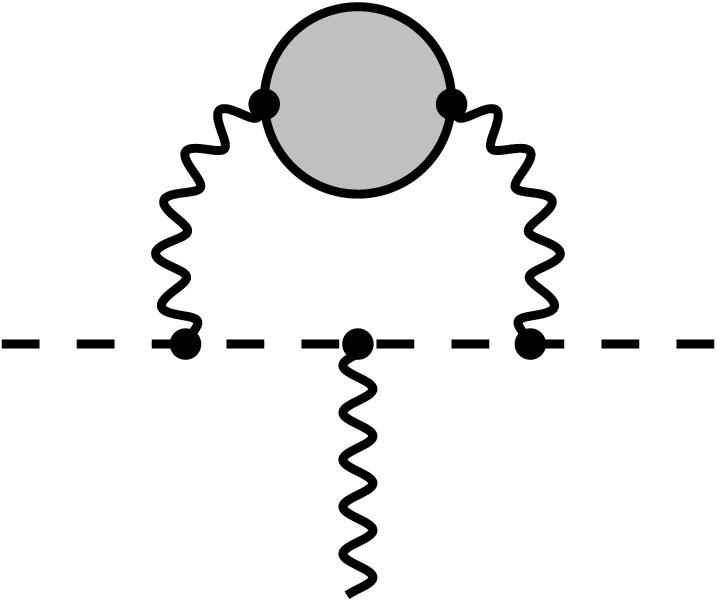}}
\put(7.5,1.9){(14)}
\put(11.5,-0.1){\includegraphics[scale=0.2,clip]{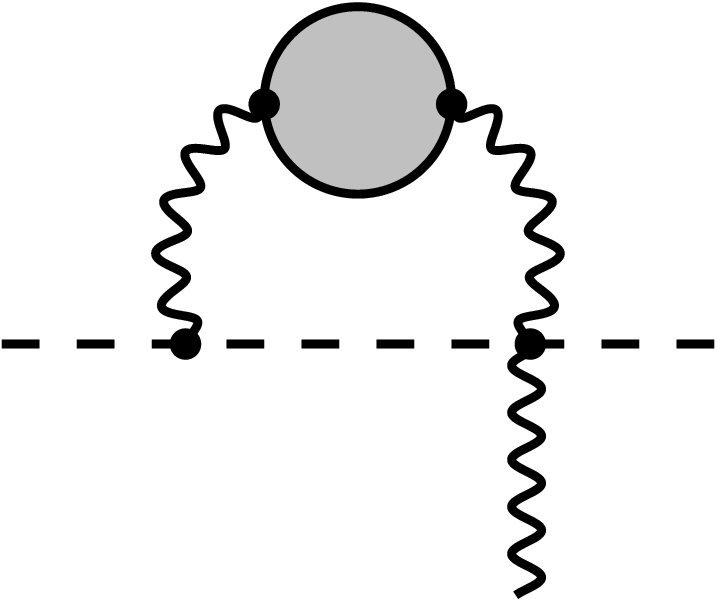}}
\put(11.5,1.9){(15)}

\end{picture}
\hspace*{2mm}
\caption{The superdiagrams giving a matter contribution to the three-point gauge-ghost vertices in the two-loop approximation. A gray disk denotes the sum of two subdiagrams presented in Fig. \ref{Figure_Polarization_Operator_Matter}.}\label{Figure_Diagrams}
\end{figure}
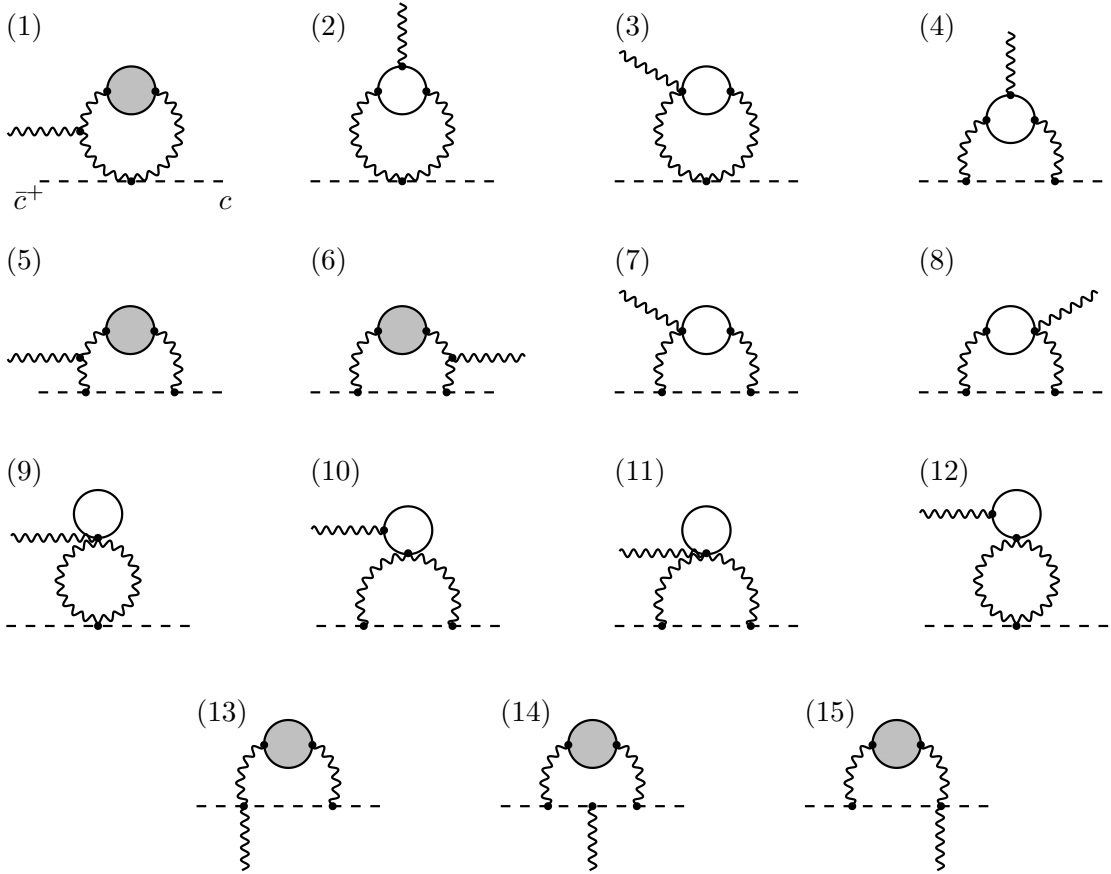

\begin{figure}[h]
\begin{picture}(0,2.5)
\put(2,0.3){\includegraphics[scale=0.18,clip]{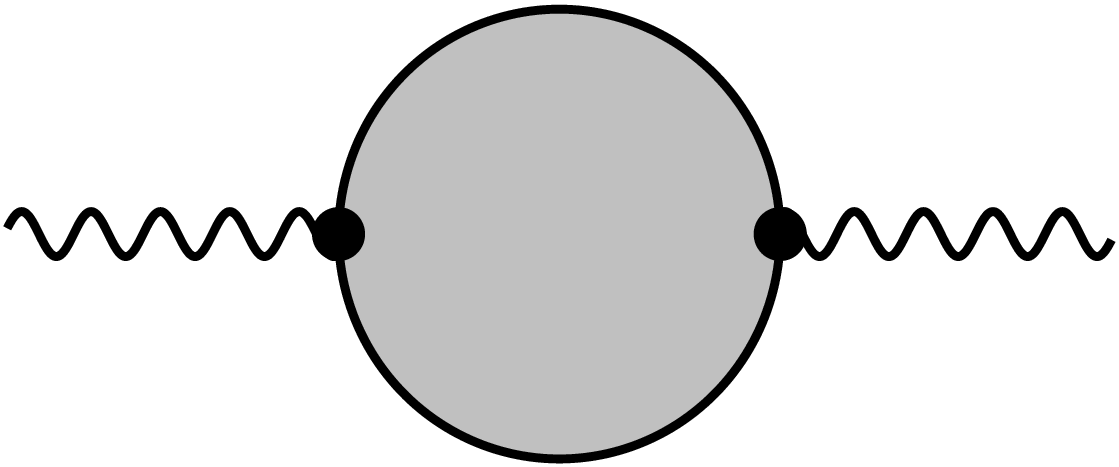}}
\put(6,1.0){$=$}
\put(7,0.3){\includegraphics[scale=0.18,clip]{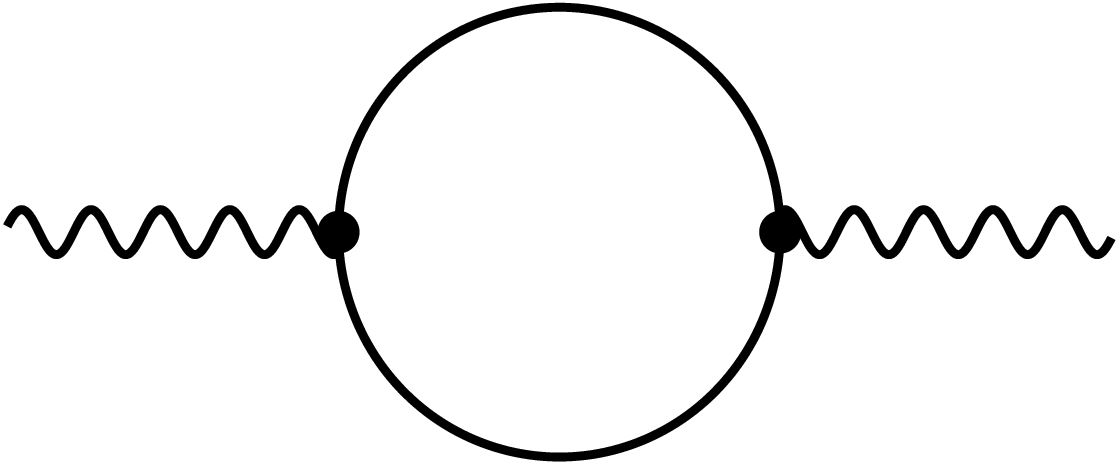}}
\put(11,1.0){$+$}
\put(11.5,0){\includegraphics[scale=0.17,clip]{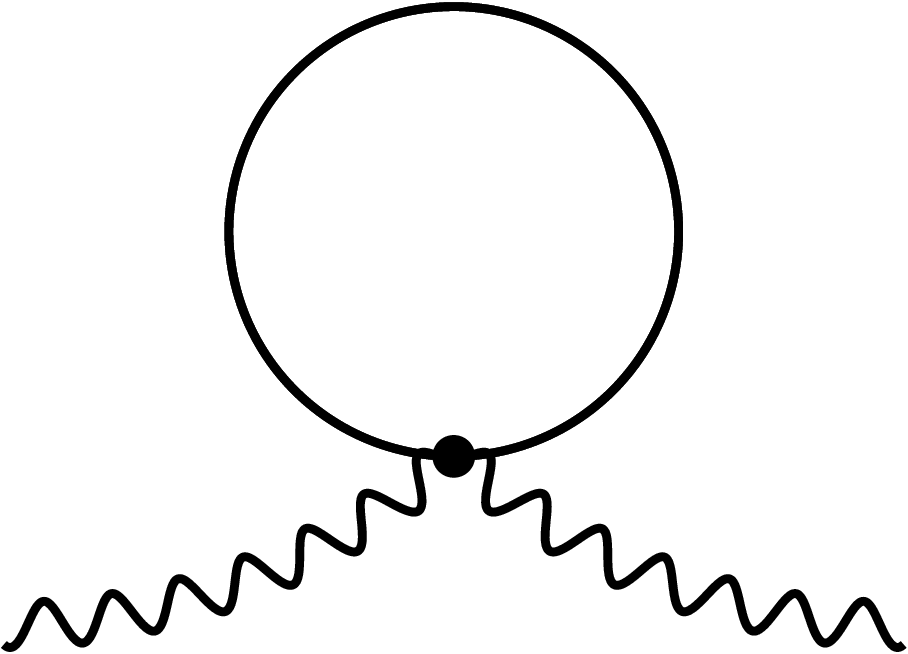}}
\end{picture}
\caption{Superdiagrams producing a matter contribution to the one-loop polarization operator of the quantum gauge superfield.}\label{Figure_Polarization_Operator_Matter}
\end{figure}

We will investigate a part of the two-loop contribution to the three-point gauge-ghost vertices coming from supergraphs containing a matter loop.\footnote{We will consider only two-loop supergraphs in which a matter loop corresponds to the usual superfields $\phi_i$ and the Pauli--Villars superfields $\Phi_i$. The supergraphs with a loop of the Pauli--Villars superfields $\varphi_{1,2,3}$ produce contributions proportional to $C_2^2$, so that it is natural to investigate them together with the two-loop supergraphs without matter loops.} They are presented in Fig. \ref{Figure_Diagrams}. Some of the diagrams presented in this figure include a gray disk which encodes a sum of two diagrams depicted in Fig. \ref{Figure_Polarization_Operator_Matter}. Actually, it corresponds to that part of the one-loop quantum gauge superfield polarization operator which comes from diagrams with a matter loop. The analytic expression for it has been found in Ref. \cite{Kazantsev:2017fdc}. In the considered massive case it is written as

\begin{equation}\label{Delta_Pi}
\Delta\Pi = -8\pi\alpha_0 T(R) \int \frac{d^4L}{(2\pi)^4}\, h(K,L),
\end{equation}

\noindent
where

\begin{eqnarray}\label{Polarization}
&&\hspace*{-9mm} h(K,L) \equiv \frac{1}{\big((K+L)^2-L^2\big)}\left(\vphantom{\frac{1}{2}}\right.  \frac{F_L^2}{2\big(L^2 F_L^2+m_0^2\big)} - \frac{F_{K+L}^2}{2\big((K+L)^2 F_{K+L}^2 + m_0^2\big)}\qquad \nonumber\\
&&\hspace*{-9mm} - \frac{m_0^2 F'_{L}}{\Lambda^2 F_L \big(L^2 F_{L}^2 + m_0^2\big)} + \frac{m_0^2 F'_{K+L}}{\Lambda^2 F_{K+L} \big((K+L)^2 F_{K+L}^2 + m_0^2\big)} - \frac{F_L^2}{2\big(L^2 F_L^2+M^2\big)}  \qquad \nonumber\\
&&\hspace*{-9mm} + \frac{F_{K+L}^2}{2\big((K+L)^2 F_{K+L}^2 + M^2\big)}   + \frac{M^2 F'_{L}}{\Lambda^2 F_L \big(L^2 F_{L}^2 + M^2\big)} - \frac{M^2 F'_{K+L}}{\Lambda^2 F_{K+L} \big((K+L)^2 F_{K+L}^2 + M^2\big)} \left.\vphantom{\frac{1}{2}}\right),
\end{eqnarray}

\noindent
and primes denote derivatives with respect to the arguments, e.g., $F'_L\equiv F'(L^2/\Lambda^2)$. The polarization operator $\Pi$ is related to the two-point Green function of the quantum gauge superfield. Taking into account that quantum corrections to this function are transversal due to the Slavnov--Taylor identities \cite{Taylor:1971ff,Slavnov:1972fg}, we can present the corresponding part of the effective action in the form

\begin{equation}\label{Dq_Definition}
\Gamma^{(2)}_V - S_{\mbox{\scriptsize gf}}^{(2)} = - \frac{1}{8\pi} \mbox{tr} \int \frac{d^4k}{(2\pi)^4}\, d^4\theta\, V(-k,\theta) \partial^2 \Pi_{1/2} V(k,\theta)\, d_q^{-1}\big(\alpha_0,\lambda_0,k^2/\Lambda^2\big).
\end{equation}

\noindent
Then the polarization operator is defined by the equation

\begin{equation}\label{Polarization_Operator}
d_q^{-1}(\alpha_0,\lambda_0,k^2/\Lambda^2) - \alpha_0^{-1} R(k^2/\Lambda^2) \equiv - \alpha_0^{-1} \Pi(\alpha_0,\lambda_0,k^2/\Lambda^2).
\end{equation}

\noindent
From the above equations it is possible to write the exact propagator of the quantum gauge superfield in terms of the function $\Pi$,

\begin{equation}\label{Effective_Propagator}
2i\left(\frac{1}{\big(R -\Pi\big)\partial^2} - \frac{1}{16\partial^4}\Big(D^2 \bar D^2 + \bar D^2 D^2\Big)\Big(\frac{\xi_0}{K} - \frac{1}{R-\Pi}\Big)\right) \delta^8_{xy} \delta^{AB},
\end{equation}

\noindent
where $\delta^8_{xy} \equiv \delta^4(x^\mu-y^\mu) \delta^4(\theta_x - \theta_y)$. Using this equation it is easy to construct expressions for various superdiagrams containing insertions of the polarization operator and, in particular, for the superdiagrams with a gray disk in Fig. \ref{Figure_Diagrams}.

To construct vertices containing ghost lines, it is necessary to take into account Eq. (\ref{Function_F}) and use the equations

\begin{eqnarray}
&& \frac{V}{1-e^{2V}} = -\frac{1}{2} + \frac{1}{2} V - \frac{1}{6} V^2 + \frac{1}{90} V^4 + O(V^6);\qquad\\
&& \frac{V}{1-e^{-2V}} = \frac{1}{2} + \frac{1}{2} V + \frac{1}{6} V^2 - \frac{1}{90} V^4 + O(V^6).
\end{eqnarray}

\noindent
Then we see that vertices with ghost legs are generated by the expression

\begin{eqnarray}\label{FP_Lower_Terms}
&& S_{\mbox{\scriptsize FP}} = \int d^4x\,d^4\theta\, \left(\frac{1}{4} c^{+A} \bar c^A + \frac{1}{4} \bar c^{+A} c^A
+ \frac{ie_0}{4} f^{ABC} (\bar c^A + \bar c^{+A}) V^B (c^C + c^{+C})  - \frac{e_0^2}{12} f^{ABC}\right.\nonumber\\
&& \times f^{CDE} (\bar c^A + \bar c^{+A}) V^B V^D (c^E - c^{+E})
- \frac{e_0^4}{180} f^{ABC} f^{CDE} f^{EFG} f^{GHI} (\bar c^A + \bar c^{+A}) V^B V^D V^F \qquad
\nonumber\\
&&\left. \times V^H (c^I - c^{+I}) -\frac{3}{4} e_0^2\, y_0\, G^{ABCD} (\bar c^A + \bar c^{+A}) V^C V^D (c^B-c^{+B}) + \ldots
\right).
\end{eqnarray}

\noindent
The term containing $y_0$ is essential even for calculating the two-loop anomalous dimension of the ghost superfields, see Ref. \cite{Kazantsev:2018kjx} for details. Therefore, it is certainly needed for calculating the two-loop contribution to the triple gauge-ghost vertices. However, it is not essential for obtaining its part proportional to $C_2 T(R)$ (which we are interested in), so that the effects of the nonlinear renormalization can be ignored in this paper. Also we see that the vertices with two ghost and three gauge lines are absent and, therefore, we need not include the corresponding superdiagrams into Fig. \ref{Figure_Diagrams}.

It is convenient to divide the superdiagrams presented in Fig. \ref{Figure_Diagrams} into three groups.

1. The superdiagrams (1), (5), and (6), in which an external gauge leg is attached to a gauge internal line.

2. The superdiagrams (13), (14), and (15) containing an insertion of the one-loop polarization operator (\ref{Delta_Pi}) and an external gauge leg attached to a ghost line.

3. The other superdiagrams (2), (3), (4), (7), (8), (9), (10), (11), and (12) in which an external gauge line is attached to a matter loop.

Let us demonstrate that the sum of superdiagrams in each of these groups is UV finite.

1. First, we consider the superdiagrams (1), (5), and (6). They include a triple gauge vertex in which (after the replacement (\ref{Formal_Substitution})) one leg corresponds to $\bar D^2 H$. The original expression for the triple gauge vertex is written as

\begin{eqnarray}
&& \Delta S_{V^3} = \frac{i e_0}{16} f^{ABC} \int d^8x\, V^A D^a V^B R(\partial^2/\Lambda^2) \bar D^2 D_a V^C \nonumber\\
&&\qquad + \frac{ie_0}{128\Lambda^2} f^{ABC} \sum\limits_{n=1}^\infty r_n \sum\limits_{\alpha=0}^{n-1} \int d^8x\, \Big(\frac{\partial^2}{\Lambda^2}\Big)^\alpha D^2 \bar D^2 D^a V^A\, V^B \Big(\frac{\partial^2}{\Lambda^2}\Big)^{n-1-\alpha} \bar D^2 D_a V^C,\qquad
\end{eqnarray}

\noindent
where $V^A$ are components of the {\it quantum} gauge superfield, $d^8x \equiv d^4x\, d^4\theta$, and the coefficients $r_n$ are defined by the equation

\begin{equation}
R(x) = 1+ \sum\limits_{n=1}^\infty r_n x^n.
\end{equation}

\noindent
After making the replacement (\ref{Formal_Substitution}) for one gauge leg, we obtain the expression

\begin{eqnarray}
&&\frac{i e_0}{16} f^{ABC} \int d^8x\,\Big(\bar D^2 H^A D^a V^B R(\partial^2/\Lambda^2) \bar D^2 D_a V^C + V^A D^a \bar D^2 H^B R(\partial^2/\Lambda^2) \bar D^2 D_a V^C \Big) \qquad\nonumber\\
&& + \frac{ie_0}{128\Lambda^2} f^{ABC} \sum\limits_{n=1}^\infty r_n \sum\limits_{\alpha=0}^{n-1} \int d^8x\, \Big(\frac{\partial^2}{\Lambda^2}\Big)^\alpha D^2 \bar D^2 D^a V^A\, \bar D^2 H^B \Big(\frac{\partial^2}{\Lambda^2}\Big)^{n-1-\alpha} \bar D^2 D_a V^C. \qquad
\end{eqnarray}

\noindent
This vertex is a part of diagrams (1), (5), and (6) which contribute to the function ${\cal S}$ defined by Eq. (\ref{Three_Point_Contribution1}). If $\bar D^2 H$ stands on the gauge external line, then the corresponding loop integrals are logarithmically divergent. Therefore, if more than two supersymmetric covariant derivatives act on the superfield $H$, the corresponding contribution is UV convergent and vanishes in the limit of the vanishing external momenta. (In the considered renormalizable theory all subdivergences are removed by the renormalization in the previous orders.) This implies that the term containing $D^a \bar D^2 H$ can be omitted. Integrating the supersymmetric covariant derivatives by parts, omitting terms with more than 2 derivatives acting on $H$, and using the identity $\bar D^2 D^2 \bar D^2 = -16\bar D^2 \partial^2$, the vertex under consideration can be rewritten as

\begin{eqnarray}
&&\frac{i e_0}{16} f^{ABC} \int d^8x\,\bigg(\bar D^2 H^A D^a V^B R(\partial^2/\Lambda^2) \bar D^2 D_a V^C  - 2 \sum\limits_{n=1}^\infty r_n \sum\limits_{\alpha=0}^{n-1} \Big(\frac{\partial^2}{\Lambda^2}\Big)^{\alpha} D^a V^A\,\qquad\nonumber\\
&& \times \bar D^2 H^B\, \Big(\frac{\partial^2}{\Lambda^2}\Big)^{n-\alpha} \bar D^2 D_a V^C\bigg).\qquad
\end{eqnarray}

\noindent
In the second term we integrate the usual space-time derivatives by parts and omit terms proportional to the external momentum $-p_\mu$ (of the superfield $H$). Then with the help of the equation

\begin{equation}
\sum\limits_{n=1}^\infty r_n n x^n = x R'(x)
\end{equation}

\noindent
we present the vertex in the limit $p \to 0$ in the form

\begin{equation}
\frac{ie_0}{16} f^{ABC} \int d^8x\, \bar D^2 H^A D^a V^B \bigg(R\Big(\frac{\partial^2}{\Lambda^2}\Big) + \frac{2\partial^2}{\Lambda^2} R'\Big(\frac{\partial^2}{\Lambda^2}\Big)\bigg) \bar D^2 D_a V^C.
\end{equation}

\noindent
Again integrating by parts (with respect to the last derivative $\bar D^2$ and the usual derivatives inside the round brackets) and omitting terms vanishing in the limit $p \to 0$ we can present this expression as a product of a symmetric (with respect to the permutation of the indices $B$ and $C$) tensor and the antisymmetric structure constants $f^{ABC}$. Certainly, such a product vanishes. Therefore, the vertex in which an external $\bar D^2 H$-leg is attached to an internal line of the quantum gauge superfield is equal to 0 in the limit of vanishing external momenta. This implies that all diagrams containing such vertices are finite. In particular, we see that the superdiagrams (1), (5), and (6) in Fig. \ref{Figure_Diagrams} are finite.

2. Superdiagrams (13), (14), and (15) also give UV finite contributions. Indeed, the ghost propagator is proportional to either

\begin{equation}\label{Ghost_Propagator}
\frac{D_x^2 \bar D_y^2}{4\partial^2} \delta^8_{xy},\qquad \mbox{or} \qquad \frac{\bar D_x^2 D_y^2}{4\partial^2} \delta^8_{xy}
\end{equation}

\noindent
depending on a sequence of the ghost vertices. As we have already discussed above, if at least one supersymmetric covariant derivative (certainly, except for $\bar D^2$ inside $\bar D^2 H$) acts on external lines, then a superdiagram evidently vanishes in the limit of the vanishing external momenta.

\begin{figure}[h]
\begin{picture}(0,4)
\put(2.5,0){\includegraphics[scale=0.35,clip]{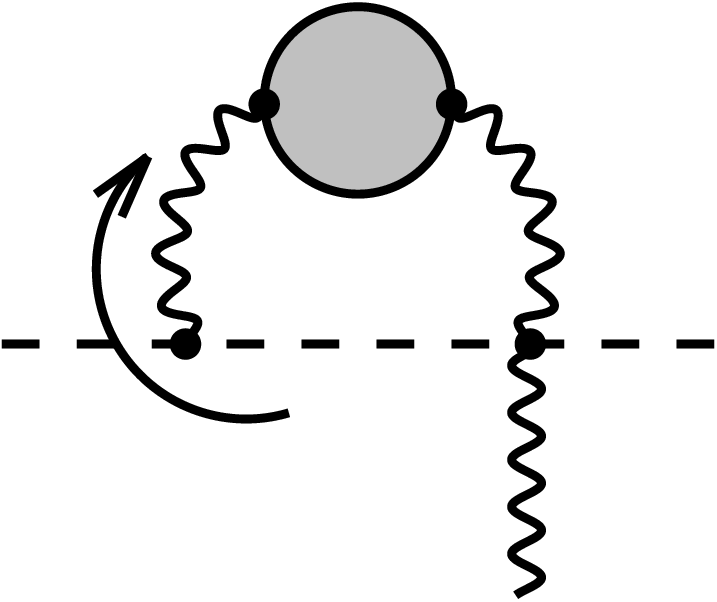}}
\put(2.5,1){$\bar c^+$} \put(6.5,1){$c$} \put(3.85,1.43){$\bm{|}$} \put(3.8,0.6){$\bar D^2$} \put(5.9,1.43){$\bm{|}$}
\put(9.5,0){\includegraphics[scale=0.35,clip]{3point_byparts.eps}}
\put(9.5,1){$\bar c^+$} \put(13.5,1){$c$} \put(12.9,1.43){$\bm{|}$} \put(10.8,0.6){$D^2$} \put(12.17,1.43){$\bm{|}$}
\end{picture}
\caption{For the superdiagrams (13), (14), and (15) in the limit of the vanishing external momenta with the help of the integration by parts one can achieve that the supersymmetric covariant derivatives act on the gauge superfield propagator.}\label{Figure_By_Parts}
\end{figure}

Let us consider a triple vertex with an external ghost leg and integrate by parts with respect to $D^2$ or $\bar D^2$ coming from ghost propagator (\ref{Ghost_Propagator}). For the superdiagram (15) this is illustrated in Fig. \ref{Figure_By_Parts}. All possible terms in which covariant derivatives act on the external ghost leg are finite. Therefore, divergences can arise only if $D^2$ or $\bar D^2$ act on the propagator of the quantum gauge superfield producing the expressions

\begin{eqnarray}
&& D^2 \left(\frac{1}{R \partial^2} - \frac{1}{16\partial^4}\Big(D^2 \bar D^2 + \bar D^2 D^2\Big)\Big(\frac{\xi_0}{K} - \frac{1}{R}\Big)\right) = \frac{\xi_0 D^2}{\partial^2 K};\nonumber\\
&& \bar D^2 \left(\frac{1}{R \partial^2} - \frac{1}{16\partial^4}\Big(D^2 \bar D^2 + \bar D^2 D^2\Big)\Big(\frac{\xi_0}{K} - \frac{1}{R}\Big)\right) = \frac{\xi_0 \bar D^2}{\partial^2 K}.\qquad
\end{eqnarray}

\noindent
The remaining derivatives $D^2$ or $\bar D^2$ act on the polarization operator, which is transversal due to Eq. (\ref{Dq_Definition}). Therefore, from the equations

\begin{equation}
D^2 \partial^2\Pi_{1/2} = 0;\qquad \bar D^2 \partial^2\Pi_{1/2} = 0
\end{equation}

\noindent
we conclude that the considered contribution vanishes. Certainly, this argumentation is valid for each of the superdiagrams (13), (14), and (15).

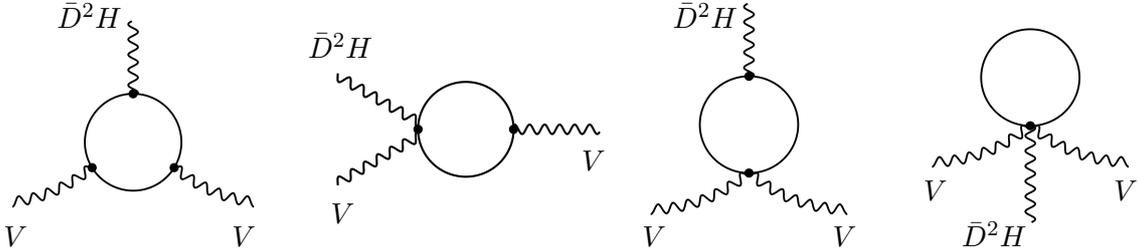
\begin{figure}[h]
\begin{picture}(0,3.6)
\put(0.6,0.5){\includegraphics[scale=0.17,clip]{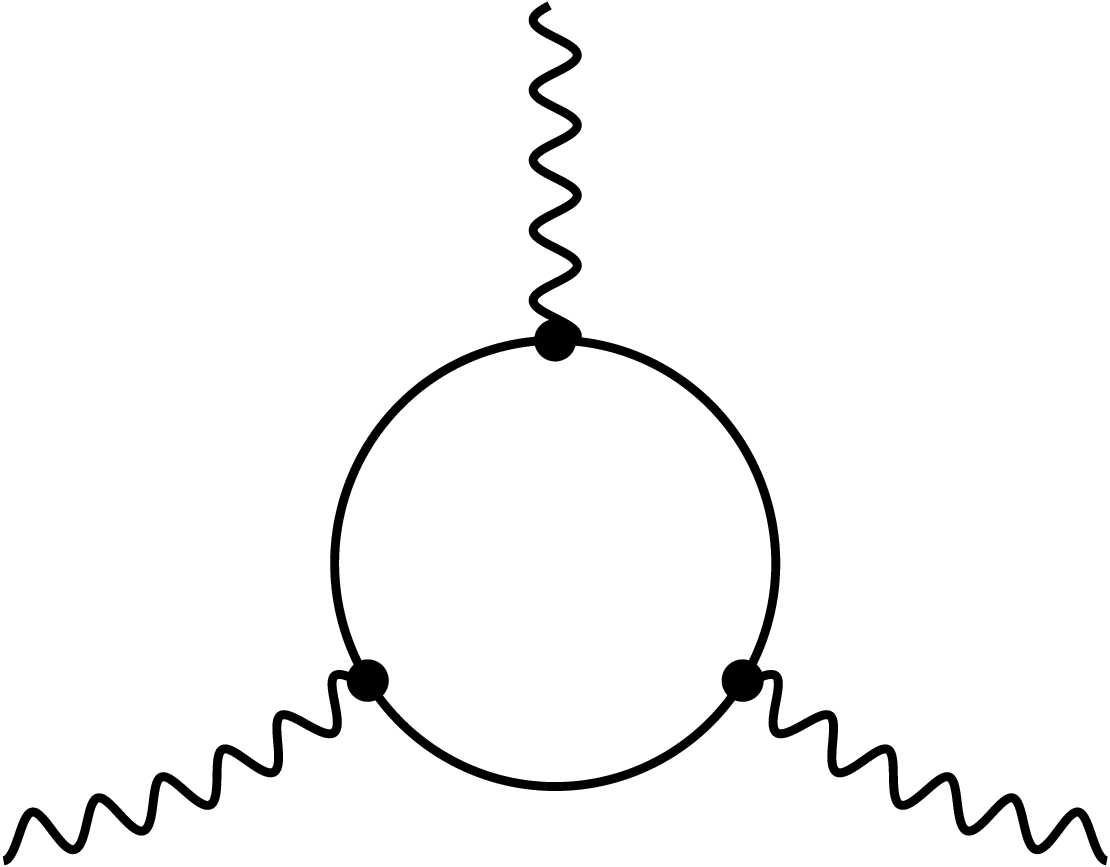}}
\put(0.5,0){$V$}\put(3.5,0){$V$} \put(1.2,2.9){$\bar D^2 H$}
\put(4.87,0.8){\includegraphics[scale=0.19,clip]{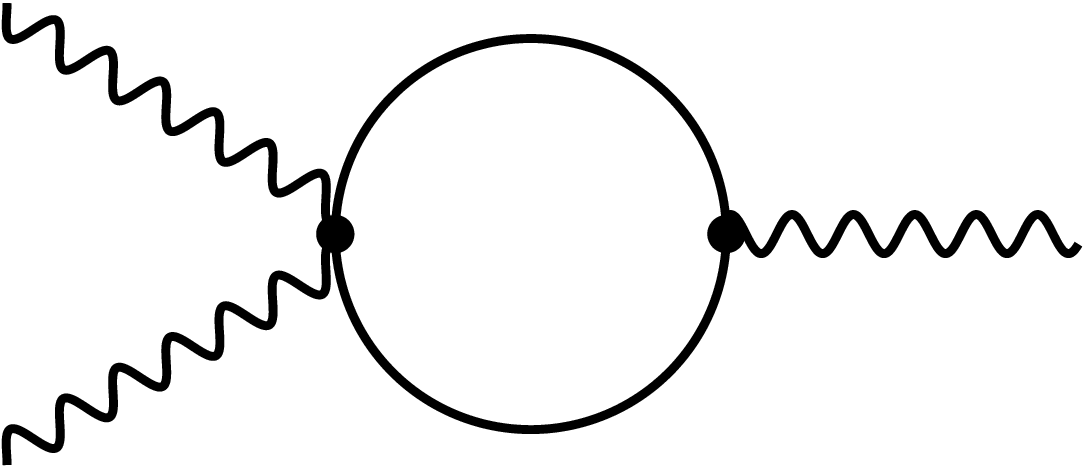}}
\put(4.8,0.3){$V$}\put(8.1,1){$V$} \put(4.5,2.5){$\bar D^2 H$}
\put(9.0,0.4){\includegraphics[scale=0.17,clip]{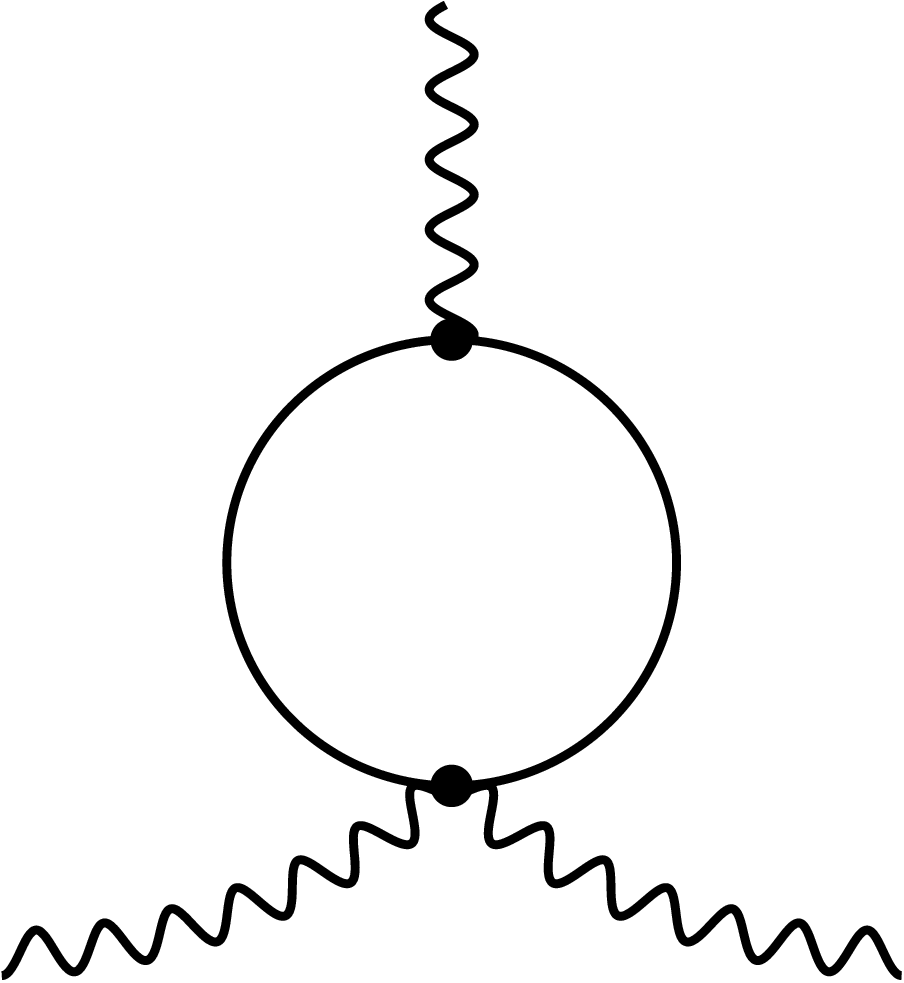}}
\put(8.9,0){$V$}\put(11.4,0){$V$} \put(9.3,2.9){$\bar D^2 H$}
\put(12.7,0.3){\includegraphics[scale=0.17,clip]{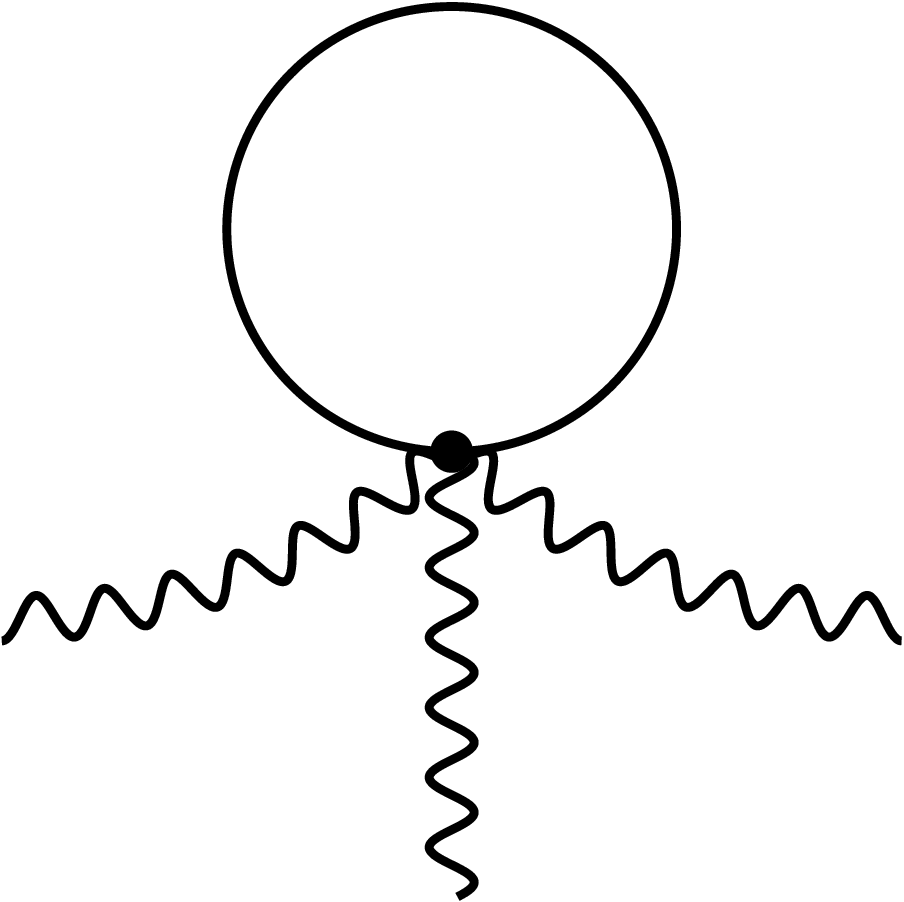}}
\put(12.6,0.6){$V$}\put(15.1,0.6){$V$} \put(13.1,0){$\bar D^2 H$}
\end{picture}
\caption{The subdiagrams obtained by attaching an external $\bar D^2 H$-leg to the matter part of the one-loop polarization operator.}\label{Figure_Polarization_Insertion}
\end{figure}

3. In the remaining supergraphs in Fig. \ref{Figure_Diagrams} an external gauge leg is attached to a matter loop. In other words, these superdiagrams contain subdiagrams presented in Fig. \ref{Figure_Polarization_Insertion}. If higher covariant derivatives are used for a regularization, then the matter propagators are given by the expressions

\begin{figure}[h]
\begin{picture}(0,2.8)
\put(6,0){\includegraphics[scale=0.2]{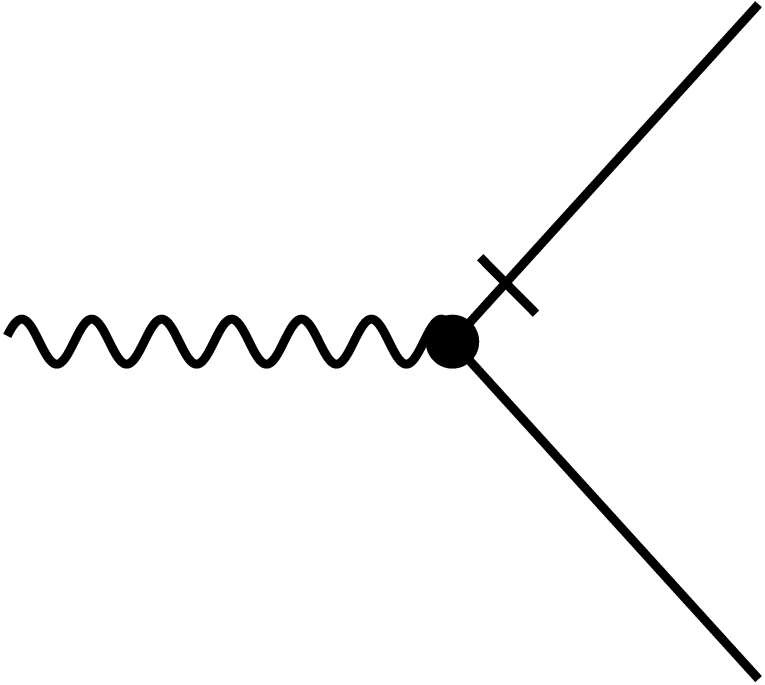}}
\put(8.7,0){$\phi^*$} \put(8.7,2){$\phi$} \put(5.5,1.1){$V$}
\end{picture}
\caption{The triple gauge-matter vertex, which in theories regularized by higher covariant derivatives is given by Eq. (\ref{Triple_Vertex}).}\label{Figure_Triple_Vertex}
\end{figure}
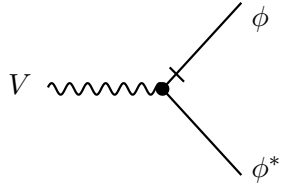

\begin{eqnarray}
&& \langle \phi^{*i}(x) \phi_j(y) \rangle = - \frac{i F}{4(\partial^2 F^2 + m_0^2)} D_x^2 \bar D_y^2 \delta^8_{xy} \delta_j^i; \qquad\nonumber\\
&& \langle \phi_i(x) \phi_j(y) \rangle = - \frac{i}{\partial^2 F^2 + m_0^2} (m_0^*)_{ij} \bar D_x^2 \delta^8_{xy};\qquad\nonumber\\
&& \langle \phi^{*i}(x) \phi^{*j}(y) \rangle = - \frac{i}{\partial^2 F^2 + m_0^2} m_0^{ij} D_x^2 \delta^8_{xy},
\end{eqnarray}

\noindent
where $x$ and $y$ are superspace points, and the regulator function $F$ (introduced in Eq. (\ref{Action_Regularized})) depends on $\partial^2$. The vertices in the regularized theory should be calculated taking into account the presence of higher derivatives in the action. For example, to obtain the simplest triple vertex depicted in Fig. \ref{Figure_Triple_Vertex}, we first present the regulator function in the second term in Eq. (\ref{Action_Regularized}) in the form $F(x) = F_0 + F_1 x + F_2 x^2 +\ldots$ Then we extract terms linear in the quantum gauge superfield from all parts of the resulting expression. They will contain the sums of the form

\begin{equation}
q^{2n} + (q+p)^2 q^{2(n-1)} + \ldots + (q+p)^{2n} = \frac{(q+p)^{2(n+1)} - q^{2(n+1)}}{(q+p)^2 - q^2}.
\end{equation}

\noindent
Calculating them using this equation and writing the result in terms of the function $F(x)$ we obtain the expression

\begin{eqnarray}\label{Triple_Vertex}
&& \frac{1}{2} \int d^4\theta \int \frac{d^4p}{(2\pi)^4} \frac{d^4q}{(2\pi)^4} \bigg(\phi^{*i}(-q-p,\theta) V_i{}^j(p,\theta)\, \phi_j(q,\theta) \frac{(q+p)^2 F_{q+p}- q^2 F_q}{(q+p)^2 - q^2}\nonumber\\
&&\qquad\qquad\qquad\qquad\qquad\qquad - \frac{1}{16} \bar D^2 \phi^{*i}(-q-p,\theta) V_i{}^j(p,\theta) D^2 \phi_j(q,\theta) \frac{F_{q+p}-F_q}{(q+p)^2 - q^2} \bigg),\qquad
\end{eqnarray}

\noindent
where

\begin{equation}
\phi_j(q,\theta) \equiv \int d^4x\, \phi_j(x,\theta) e^{i q_\alpha x^\alpha}\quad \mbox{etc.}
\end{equation}

\noindent
Vertices with a larger number of gauge legs can be found similarly, but the calculations and the resulting expressions are much more complicated.

Using the explicit expressions for vertices and propagators we find the contributions of the superdiagrams presented in Fig. \ref{Figure_Polarization_Insertion}. Investigating them it is necessary to take into account that the considered theory satisfies Eqs. (\ref{Anomalies}) and (\ref{Anomalies_PV}). After a rather non-trivial calculation we have obtained that in the limit $p\to 0$ their sum is proportional to

\begin{eqnarray}\label{Triple_Subdiagrams}
&&\hspace*{-6mm} e_0^3 f^{ABC} T(R) \int d^4\theta\, \bar D^2 H^A(0,\theta) \int \frac{d^4k}{(2\pi)^4}  [\bar D_{\dot a}, D_b] V^B(k,\theta) V^C(-k,\theta) (\gamma^\mu)^{\dot a b} \int \frac{d^4q}{(2\pi)^4} \frac{(2q+k)_\mu}{(q+k)^2-q^2}\nonumber\\
&&\hspace*{-6mm} \times F_{q+k} F_q \bigg(\frac{1}{q^2 F_q^2 - m_0^2} - \frac{1}{q^2 F_q^2 - M^2} - \frac{1}{(q+k)^2 F_{q+k}^2 - m_0^2} + \frac{1}{(q+k)^2 F_{q+k}^2 - M^2}\bigg).
\end{eqnarray}

\noindent
(Certainly, here we omitted the integral over $d^4p$, because we are interested only in the form of the momentum integral in the limit $p \to 0$, and a numerical coefficient, which will not be essential below.) Evidently, after the Wick rotation the integral in this equation,

\begin{eqnarray}
&& I_\mu \equiv \int \frac{d^4Q}{(2\pi)^4} \frac{(2Q+K)_\mu F_{Q+K} F_Q}{(Q+K)^2-Q^2}\bigg(\frac{1}{Q^2 F_Q^2 +m_0^2} - \frac{1}{Q^2 F_Q^2 + M^2}\nonumber\\
&&\qquad\qquad\qquad\qquad\qquad\qquad\qquad - \frac{1}{(Q+K)^2 F_{Q+K}^2+m_0^2} + \frac{1}{(Q+K)^2 F_{Q+K}^2+M^2}\bigg),\qquad
\end{eqnarray}

\noindent
will be proportional to $K_\mu$. Taking into account that

\begin{equation}
K^\nu (2Q+K)_\nu = (Q+K)^2 - Q^2,
\end{equation}

\noindent
we see that it can equivalently be presented in the form

\begin{eqnarray}\label{I_Final}
&& I_\mu = \frac{K_\mu K^\nu}{K^2} I_\nu = \frac{K_\mu}{K^2} \int \frac{d^4Q}{(2\pi)^4} F_{Q+K} F_Q \bigg(\frac{1}{Q^2 F_Q^2 +m_0^2} - \frac{1}{Q^2 F_Q^2 + M^2}\bigg)\nonumber\\
&& - \frac{K_\mu}{K^2} \int \frac{d^4Q}{(2\pi)^4} F_{Q+K} F_Q \bigg(\frac{1}{(Q+K)^2 F_{Q+K}^2+m_0^2} - \frac{1}{(Q+K)^2 F_{Q+K}^2+M^2}\bigg) = 0.\qquad
\end{eqnarray}

\noindent
The last equality is obtained after the change of the integration variable $Q_\mu \to -Q_\mu-K_\mu$ in the second integral if we take into account that $F_{-Q} = F_Q$ and $F_{-Q-K} = F_{Q+K}$. This change of variable is possible, because the above integral is only logarithmically divergent due to the contribution of the Pauli--Villars superfields.

Due to Eqs. (\ref{Triple_Subdiagrams}) and (\ref{I_Final}) the sums of the superdiagrams (2), (3), (9), (12) and (4), (7), (8), (10), (11) turn out to be UV finite.

Thus, we conclude that in the limit of the vanishing external momenta the sum of supergraphs presented in Fig. \ref{Figure_Diagrams} vanishes. This, in turn, implies that for non-vanishing external momenta this sum (expressed in terms of the renormalized couplings) is given by finite integrals in the exact agreement with the non-renormalization theorem for the triple gauge-ghost vertices.

\section{Conclusion}
\hspace*{\parindent}

In this paper we have demonstrated that a part of the two-loop contribution to the three-point gauge-ghost vertices proportional to $C_2 T(R)$, which comes from superdiagrams containing a matter loop, is finite in the UV region. These vertices have two external ghost legs and one leg of the quantum gauge superfield. The calculation has been done with the help of the higher covariant derivative regularization in the limit of the vanishing external momenta for a general $\xi$-gauge. The result completely agrees with the general statement derived in Ref. \cite{Stepanyantz:2016gtk}, according to which the considered vertices are finite in all orders of the perturbation theory. Unlike some similar previous results \cite{Dudal:2002pq,Capri:2014jqa}, it was proved in the case of using the superfield formulation of ${\cal N}=1$ supersymmetric gauge theories for a general $\xi$-gauge. In the supersymmetric case this statement turned out to be a very important step for the perturbative derivation of the non-Abelian NSVZ $\beta$-function made in Refs. \cite{Stepanyantz:2016gtk,Stepanyantz:2019ihw,Stepanyantz:2020uke}. That is why the result of this paper can be treated as a check of a certain part of this proof. However, it should be noted that the total two-loop contribution to the three-point gauge-ghost vertices also includes terms proportional to $C_2^2$, which were not considered in this paper. We hope to analyze them in the forthcoming publications.

\vspace{5mm}

\section*{Acknowledgments}
\hspace*{\parindent}

The authors are very grateful to A.~L.~Kataev for valuable discussions.

This work was supported by Foundation for Advancement of Theoretical Physics and Mathematics ``BASIS'', grants  No. 19-1-1-45-5 (M.K.), No. 18-2-6-159-1 (N.M.), No. 18-2-6-158-1 (S.N.), No. 19-1-1-45-3 (I.S.), and No. 19-1-1-45-1 (K.S.).

\end{document}